\documentclass[letterpaper,prl,twocolumn,showpacs,superscriptaddress]{revtex4}

\usepackage{graphicx}
\usepackage{amsmath}

\newcommand{\eg} {{e.g., }}
\newcommand{\ein} {\epsilon_{\rm in}}
\newcommand{\eout} {\epsilon_{\rm out}}
\newcommand{\einout} {\epsilon_{\rm in,out}}
\newcommand{\epps} {\epsilon_{\rm PS}}

\newcommand{\ie} {{i.e., }}
\newcommand{\pbar} {\bar{p}}

\newcommand{\tenG} {{\bf G}}
\newcommand{\tenI} {{\bf I}}
\newcommand{\vecr} {{\bf r}}
\newcommand{\vecR} {{\bf R}}
\newcommand{\vecz} {{\bf z}}

\newcommand{\zhat} {\hat{\vecz}}

\begin{document}

\title
{Enhanced dispersion interaction in confined geometry}

\author{Michal Marcovitch}
\affiliation{School of Physics \& Astronomy}

\author{Haim Diamant}
\email{hdiamant@tau.ac.il}
\affiliation{School of Chemistry\\ 
Raymond and Beverly Sackler Faculty
of Exact Sciences, Tel Aviv University,
Tel Aviv 69978, Israel}

\date{October 2, 2005}

\begin{abstract}
  The dispersion interaction between two point-like particles confined
  in a dielectric slab between two plates of another dielectric medium
  is studied within a continuum (Lifshitz) theory.  The retarded
  (Casimir-Polder) interaction at large interparticle distances is
  found to be strongly enhanced as the mismatch between the dielectric
  permittivities of the two media is increased.  The large-distance
  interaction is multiplied due to confinement by a factor of
  $(33\gamma^{5/2}+13\gamma^{-3/2})/46$ at zero temperature, and by
  $(5\gamma^2+\gamma^{-2})/6$ at finite temperature,
  $\gamma=\ein(0)/\eout(0)$ being the ratio between the static
  dielectric permittivities of the inner and outer media.  This
  confinement-induced amplification of the dispersion interaction can
  reach several orders of magnitude.
\end{abstract}


\maketitle

The dispersion interaction acts between any two polarizable objects,
thus being one of the most ubiquitous interactions in nature
\cite{MN,Parsegian}. It plays a central role in numerous
phenomena in chemical physics and materials science, including
gas--liquid condensation, capillarity \cite{Widom}, inter-surface
interactions \cite{Israelachvili}, and colloid stability
\cite{Russel}.

The dispersion interaction is a quantum fluctuation-induced coupling
between two polarizable particles mediated by the electromagnetic
field. In an unconfined system at zero temperature there is a single
length scale with which the interparticle distance $R$ is to be
compared, \ie the characteristic wavelength $\lambda_0$ of photon
absorption by the particles, typically in the ultraviolet to visible
range. London's calculation \cite{London}, valid in the nonretarded
limit $R\ll\lambda_0$, yields the potential $U(R) =
-[(3\hbar/\pi)\int_0^\infty d\xi \alpha^2(i\xi)]R^{-6}$, where
$\alpha(\omega)$ is the frequency-dependent polarizability of the
particles.
Casimir and Polder \cite{CP} recast the problem in
quantum-electrodynamic terms, whereby the interaction arises from the
effect of the particles on the zero-point modes of the electromagnetic
field. For $R\ll\lambda_0$ the Casimir-Polder result coincides with
London's, yet in the retarded limit, $R\gg\lambda_0$, the interaction
decays as $R^{-7}$, $U(R)=-[(23/(4\pi))\hbar c\alpha^2(0)]R^{-7}$
\cite{ft_static}.  As a result, the dispersion interaction between two
particles in the retarded regime (typically $R>0.1$ $\mu$m) is
extremely weak and has not been directly observed.
(Particle--surface and surface--surface interactions across such
micron-scale distances are much stronger and were successfully
measured in the 1990s \cite{Sukenik,Lamoreaux}.)  In the current
Letter we demonstrate that this weak particle--particle interaction
can be dramatically amplified in confined geometries.

At a finite temperature $T$ another length scale appears, \ie the
thermal wavelength $\lambda_T = \hbar c/T\simeq 7.6$ $\mu$m at room
temperature. (The Boltzmann constant is set hereafter to unity.)  As
was shown in Ref.\ \cite{Ninham9899}, retardation and
finite-temperature effects are intertwined. For $R\gg\lambda_T$, the
Helmholtz free energy of interaction returns to a $R^{-6}$ dependence,
$F(R)=-3T\alpha^2(0)R^{-6}$.

The theory of dispersion interactions was extended by Lifshitz {\it et
al}.\ to the case where the interacting objects as well as the
intervening space are continuous media \cite{Lifshitz,Pitaevskii}.  In
this continuum theory the material response to electromagnetic fields
is assumed to be fully captured by the complex, frequency-dependent
dielectric permittivity $\epsilon(\omega)$ \cite{ft_frequency}.  The
current work is based on such a continuum approach. Despite the strong
underlying assumption (after all, the media themselves consist of
discrete polarizable particles), the Lifshitz theory has been widely
used and experimentally corroborated \cite{Israelachvili}.  It is
expected to yield valid results so long as the distance between the
two particles is much larger than the intermolecular distances in the
materials.

In various circumstances particles are spatially confined, \eg in
porous media, micro-cavities, biological constrictions or nanofluidic
devices. Such confinement introduces a new length scale, the
separation $h$ between the bounding surfaces. The dispersion
interaction between a single particle and confining surfaces has been
extensively studied in the context of cavity QED
\cite{Sukenik,cavity}.  The effect of confinement on the interaction
between two particles, however, has been only partially addressed.
Confinement by two metallic plates (\ie the limit where the
permittivity of the outer medium $\eout\rightarrow\infty$) was found
to drastically affect the interaction between two point-like particles
in vacuum at zero \cite{Ninham73} and nonzero \cite{Ninham01}
temperatures. The nonretarded interaction between particles confined
by two dielectric plates at $T=0$ was addressed within a single-image
approximation in Ref.\ \cite{Silby}. We follow the lines of Mahanty,
Ninham, Bostr\"om and Longdell \cite{Ninham73,Ninham01} and extend
their theory to the general and more practical case of arbitrary
permittivities of both inner and outer media, discovering a dramatic
enhancement for experimentally relevant values of $\ein$ and $\eout$.

\begin{figure}[tbh]
\centerline{\resizebox{0.37\textwidth}{!}
{\includegraphics{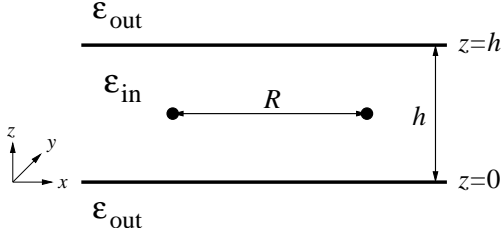}}}
\caption[]{Schematic view of the system and its parameters.}
\label{fig_system}
\end{figure}

The system under consideration is schematically shown in Fig.\ 
\ref{fig_system}. Two point-like, isotropic particles of
polarizability $\alpha(\omega)$ are embedded in a slab of thickness
$h$ and dielectric permittivity $\ein(\omega)$. The slab is bounded by
two semi-infinite media of dielectric permittivity $\eout(\omega)$.
We use cylindrical coordinates, $\vecr=(\rho,\varphi,z)$, the $\zhat$
axis taken perpendicular to the bounding surfaces. For simplicity we
specialize to the symmetric case where the particles lie on the slab
midplane, $z=h/2$, connected by the vector $\vecR=(R,0,0)$.  All
materials are assumed nonmagnetic.

We employ the semiclassical scheme introduced by Mahanty and Ninham,
which accurately reproduces the London and Casimir-Polder results
\cite{Ninham72}. In this theory the energy of interaction at $T=0$ is
given by \cite{MN}
\begin{eqnarray}
  &&T=0:\ \ U(\vecR) = -8\pi\hbar \int_0^\infty d\xi 
  \frac{\xi^4}{c^4}\alpha^2(i\xi) 
\nonumber\\
  &&\times {\rm Tr} [\tenG(\vecR,i\xi) \tenG(-\vecR,i\xi)],
\label{U1}
\end{eqnarray}
and the Helmholtz free energy at finite $T$ by
\begin{eqnarray}
  &&T>0:\ \ F(\vecR) = -(4\pi)^2 T \sideset{}{'}\sum_{n=0}^\infty
  \alpha^2(i\xi_n) (\xi_n^4/c^4) 
\nonumber\\
  &&\times{\rm Tr} [\tenG(\vecR,i\xi_n) \tenG(-\vecR,i\xi_n)],\ \  
  \xi_n = (2\pi c/\lambda_T)n,
\label{F1}
\end{eqnarray}
the prime indicating that the $n=0$ term is multiplied by $1/2$. In
Eqs.\ (\ref{U1}) and (\ref{F1}) $\vecR=\vecr-\vecr'$, and
$\tenG(\vecr,\vecr',\omega)$ is the dyadic Green tensor of the
electric-field wave equation,
\begin{equation}
  \nabla\times\nabla\times\tenG - \epsilon_m(\omega)(\omega^2/c^2)
  \tenG = \tenI\delta(\vecr-\vecr'),
\label{wave}
\end{equation}
where $\tenI$ is the identity tensor and $\epsilon_m=\ein$ or $\eout$
depending on whether $\vecr$ lies in the inner or outer medium. (The
position $\vecr'$ is taken inside the slab.) The boundary conditions
in the current case are continuity across the bounding surfaces of the
tangential components of both the electric and magnetic fields. This
imposes continuity on $\zhat\times\tenG$ and
$\zhat\times\nabla\times\tenG$ across $z=0$ and $z=h$.

Thus, this scheme reduces the problem to finding the Green tensor
$\tenG(\vecr,\vecr',\omega)$ of Eq.\ (\ref{wave}) with the
aforementioned boundary conditions. The derivation is technically
complicated and can be found in Refs.\ \cite{Tai,msc}. In the
symmetric case of interest, $\vecr'=(0,0,h/2)$ and $\vecr=(R,0,h/2)$,
the tensor becomes diagonal,
\begin{eqnarray}
  && \frac{\xi^2}{c^2} G_{ij}(R,i\xi) = 
  \frac{1}{4\pi\ein(i\xi)R^{3}}
  \int_0^\infty dx g_i(R,i\xi,x)\delta_{ij}
\nonumber\\
  && g_\rho = [q^2x (e^v - t) (e^v - u) J_0(x) - 
  (x^2(e^{2v} + tu)
\nonumber\\
  && \ \ \ - e^v (p^2 + q^2)(t+u))J_1(x)] / [q(e^v-t)(e^v + u)]
\nonumber\\
  && g_\varphi = [p^2x (e^v + t) (e^v + u) J_0(x) + 
  (x^2(e^{2v} + tu)
\nonumber\\
  && \ \ \ - e^v (p^2 + q^2)(t+u))J_1(x)] / [q(e^v-t)(e^v + u)]
\nonumber\\
  && g_z = -x^3 (e^v+u)J_0(x) / [q(e^v - u)],
\label{green}
\end{eqnarray}
where $J_k(x)$ are Bessel functions, and the following abbreviations
have been used:
$\gamma = \ein/\eout$,
$p = R\xi\ein^{1/2}/c$,
$q = (x^2+p^2)^{1/2}$,
$s = (x^2+p^2/\gamma)^{1/2}$,
$t = (q-s)/(q+s)$,
$u = (q-\gamma s)/(q+\gamma s)$,
and $v = hq/R$.
Given $\einout(i\xi)$ and $\alpha(i\xi)$, one can substitute Eq.\
(\ref{green}) in Eq.\ (\ref{U1}) or (\ref{F1}) and calculate
numerically the interaction potential.

It is instructive, however, to first analyze the interaction in
several asymptotic limits. We begin with the small-distance limit,
recovering the known results for unconfined particles. For $R\ll h$
the expressions for $g_i$ are expanded to leading order in large $v$,
whereupon the integration in Eq.\ (\ref{green}) can be carried out
analytically.  Substituting the result in Eq.\ (\ref{U1}), we get
$U(R) = -(\hbar/\pi)R^{-6}\int_0^\infty d\xi e^{-2p}(3+6p+5p^2+2p^3+p^4)
\alpha^2(i\xi)/\ein^2(i\xi)$.
In the nonretarded limit, $R\ll\lambda_0$, we take the leading
order in small $p$, for which the London result is recovered,
\begin{equation}
  T=0,\ R\ll h,\lambda_0:\  
  U(R) = -\frac{3\hbar}{\pi R^6} \int_0^\infty d\xi 
  \frac{\alpha^2(i\xi)}{\ein^2(i\xi)},
\label{London}
\end{equation}
with the appropriate correction due to the fact that the particles
are not in vacuum ($\ein\neq 1$).  In the retarded limit,
$R\gg\lambda_0$, the integral over $\xi$ is dominated by small
(static) $\xi$. An asymptotic analysis then recovers the
Casimir-Polder result with the correction for $\ein\neq 1$,
\begin{equation}
  T=0,\ \lambda_0\ll R\ll h:\ 
  U(R) = -\frac{23\hbar c\alpha^2(0)}{4\pi\ein^{5/2}(0) R^{7}}.
\label{CP}
\end{equation}

Similarly, substitution of the asymptotic Green tensor for $R\ll h$ in
Eq.\ (\ref{F1}) yields
$F(R) = -2T R^{-6} \sum_{n=0}^{'\infty}
e^{-2p_n}(3+6p_n+5p_n^2+2p_n^3+p_n^4)
\alpha^2(i\xi_n)/\ein^2(i\xi_n)$,
where $p_n = R\xi_n\ein^{1/2}/c$. Subsequently taking the limit
$R\ll\lambda_T$, we expand to leading order in small $p_n$
\cite{ft_Tto0} and recover the known result for the nonretarded
interaction at finite temperature \cite{McLachlan},
\begin{equation}
  T>0,\ R\ll h,\lambda_T:\ 
  F(R) = -\frac{6T}{R^6} \sideset{}{'}\sum_{n=0}^\infty \frac{\alpha^2(i\xi_n)}
  {\ein^2(i\xi_n)}.
\label{McLachlan}
\end{equation}
In the other limit, $R\gg\lambda_T$, the sum is dominated by the $n=0$
term, leading to the known retarded interaction at finite temperature
\cite{Ninham9899},
\begin{equation}
  T>0,\ \lambda_T\ll R\ll h:\ 
  F(R) = -\frac{3T\alpha^2(0)}{\ein^2(0)R^6}.
\label{Ninham99}
\end{equation}

We now turn to the more interesting asymptotic limit of large $R$
where confinement sets in. In the limit $R\gg h$ the expressions for
$g_i$ in Eq.\ (\ref{green}) are expanded to leading order in small
$v$, and the integration in Eq.\ (\ref{green}) is performed
analytically. The result is substituted in Eq.\ (\ref{U1}) to yield
$U(R) = -(\hbar/(2\pi))R^{-6}\int_0^\infty d\xi e^{-2\pbar}
[(5+\gamma^{-4}) + (10+2\gamma^{-4})\pbar + (7+3\gamma^{-4})\pbar^2 
+ (2+2\gamma^{-4})\pbar^3 + (1+\gamma^{-4})\pbar^4] 
\alpha^2(i\xi)/\eout^2(i\xi)$,
where $\pbar=\gamma^{-1/2}p$. In the nonretarded regime, this expression
is expanded to leading order in small $\pbar$, leading to
\begin{eqnarray}
\label{London_conf}
  && T=0,\ h\ll R\ll\lambda_0:\ \\
  && U(R) = -\frac{3\hbar}{\pi R^6} \int_0^\infty d\xi 
  \frac {\alpha^2(i\xi)}{\ein^2(i\xi)}
  \frac{5\gamma^2(i\xi) + \gamma^{-2}(i\xi)}{6}.\nonumber
\end{eqnarray}
%
%
%
Equation (\ref{London_conf}) gives the nonretarded interaction under
confinement. It reduces to the London result, Eq.\ (\ref{London}), for
$\gamma=\ein/\eout=1$. This result has a rather restricted validity as
it requires that $h$ be much smaller than $\lambda_0$. (A detailed
discussion of the confined nonretarded interaction will be given
elsewhere.) In the retarded regime the integral is dominated by small
$\xi$, leading to
\begin{eqnarray}
  &&T=0,\ R\gg h,\lambda_0:\ \nonumber\\
  &&U(R) = -\frac{23\hbar c\alpha^2(0)}{4\pi\ein^{5/2}(0) R^{7}}
  \frac{33\gamma^{5/2} + 13\gamma^{-3/2}}{46},
\label{CP_conf}
\end{eqnarray}
where we have written $\gamma(0)=\gamma$ for brevity. This 
retarded interaction under confinement converges to the
Casimir-Polder expression, Eq.\ (\ref{CP}), for $\gamma=1$.

At finite temperature we substitute the asymptotic Green tensor for
$R\gg h$ in Eq.\ (\ref{F1}), resulting in
$F(R) = -TR^{-6} \sum_{n=0}^{'\infty} e^{-2\pbar_n}
[(5+\gamma^{-4}) + (10+2\gamma^{-4})\pbar_n + (7+3\gamma^{-4})\pbar_n^2 
+ (2+2\gamma^{-4})\pbar_n^3 + (1+\gamma^{-4})\pbar_n^4] 
\alpha^2(i\xi_n)/\eout^2(i\xi_n)$,
where $\pbar_n=R\xi_n\eout^{1/2}/c$. For $R\ll\lambda_T$ we take the
leading order in small $\pbar_n$ \cite{ft_Tto0} and get
\begin{eqnarray}
  &&T>0,\ h\ll R\ll \lambda_T:\ 
\label{McLachlan_conf}\\
  &&F(R) = -\frac{6T}{R^6} \sideset{}{'}\sum_{n=0}^\infty 
  \frac{\alpha^2(i\xi_n)}{\ein^2(i\xi_n)}
  \frac{5\gamma^2(i\xi_n)+\gamma^{-2}(i\xi_n)}{6}. \nonumber
\end{eqnarray}
%
%
This is the extension of Eq.\ (\ref{McLachlan}) to the confined case.
For $R\gg\lambda_T$, the $n=0$ term dominates the sum, yielding
\begin{equation}
  T>0,\ R\gg h,\lambda_T:\ 
  F(R) = -\frac{3T\alpha^2(0)}{\ein^2(0)R^6}
  \frac{5\gamma^2+\gamma^{-2}}{6},
\label{Ninham99_conf}
\end{equation}
which extends Eq.\ (\ref{Ninham99}) to the confined geometry.

Equations (\ref{CP_conf}) and (\ref{Ninham99_conf}) are our central
results. They account for the large-distance, retarded dispersion
interaction between the confined particles at zero and finite
temperature, respectively.  Comparing with Eqs.\ (\ref{CP}) and
(\ref{Ninham99}), we see that the confinement is manifest as a factor
dependent on the ratio $\gamma(0)$ between the static dielectric
permittivities of the inner and outer media. This factor can be as
small as $0.78$ [Eq.\ (\ref{CP_conf})] or $0.75$ [Eq.\
(\ref{Ninham99_conf})], but increases indefinitely with the
permittivity mismatch. The divergence of the interaction energy for
$\gamma\rightarrow 0$ or $\infty$ is obviously unphysical. Although
Eqs.\ (\ref{CP_conf}) and (\ref{Ninham99_conf}) are asymptotically
correct for any finite mismatch, as $\gamma$ becomes increasingly
large or small one must go to ever larger interparticle distances for
these asymptotes to hold. Ultimately, in the limits $\gamma\rightarrow
0,\infty$ their range of validity disappears, and the large-distance
interaction obeys a different power law
\cite{Ninham73,Ninham01}. (Detailed analysis of this behavior will be
given elsewhere.) The main point, however, is that the amplification
factor can be very large for reasonable values of $\gamma$. For
example, for particles embedded in a polar liquid [$\ein(0)=80$] which
is confined by glass plates [$\eout(0)=4$] at room temperature, one
gets an amplification factor of about 300. If the outer medium is a
gas (a free-standing film, $\eout=1$), the factor increases to about
5000 \cite{ft_salt}.

\begin{figure}[tbh]
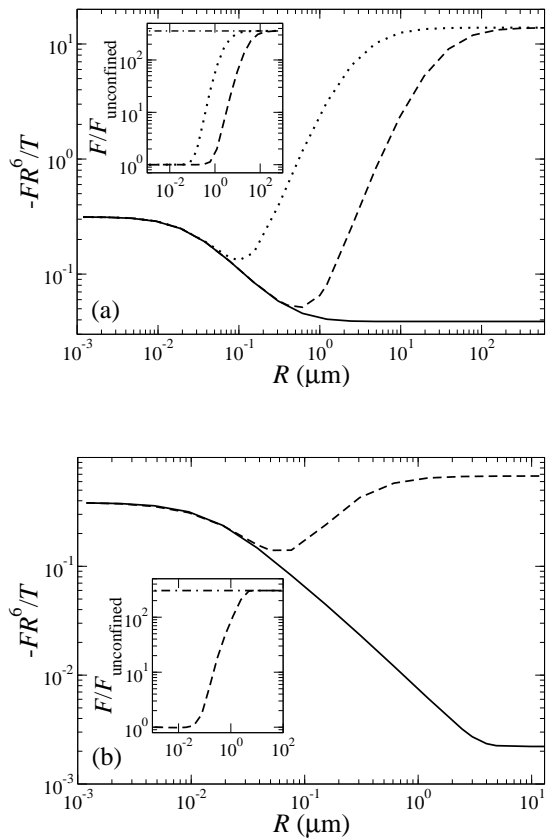

\centerline{\resizebox{0.4\textwidth}{!}
{\includegraphics{fig2a.eps}}}
\vspace{0.85cm}
\centerline{\resizebox{0.4\textwidth}{!}
{\includegraphics{fig2b.eps}}}
\caption[]{Potentials of interaction as obtained from numerical integration
  of Eqs.\ (\ref{F1}) and (\ref{green}). The interaction free energy
  is scaled by $-T/R^6$. (a) Polystyrene particles in water between
  two glass plates at $T=300$K. Solid, dashed, and dotted curves
  correspond, respectively, to an unconfined system, $h=1$ $\mu$m, and
  $h=0.1$ $\mu$m.  (b) Polystyrene particles in a hydrocarbon slab
  embedded in water at $T=300$K. Solid and dashed curves correspond to
  an unconfined system and $h=3$ nm, respectively. Insets in both
  panels present the ratio between the confined and unconfined
  potentials, the dash-dotted lines showing the asymptotic
  amplification factor of Eq.\ (\ref{Ninham99_conf}).}
\label{fig_numer}
\end{figure}

Finally, we present results from numerical integration of Eqs.\
(\ref{F1}) and (\ref{green}) for two examples of practical
interest. This requires expressions for $\epsilon(i\xi)$ of the various
media, for which we use the Ninham-Parsegian representation
\cite{Parsegian,Bergstrom}, an empirical fit based on electromagnetic
absorption spectra of the materials.
In the first example two polystyrene particles are confined in a slab
of water between two glass plates at room temperature. The function
$\ein(i\xi)$ for water is found in Refs.\ \cite{Parsegian,Russel} and
that for silica glass, $\eout(i\xi)$, in Ref.\ \cite{Bergstrom}. For
$\alpha(i\xi)$ we took the excess Clausius-Mossotti polarizability
\cite{Israelachvili} of a polystyrene sphere,
$\alpha(i\xi)/V=[3\ein(i\xi)/(4\pi)]
[\epps(i\xi)-\ein(i\xi)]/[\epps(i\xi)+2\ein(i\xi)]$, $V$ being the
particle volume and $\epps(i\xi)$ the permittivity of polystyrene,
found in Refs.\ \cite{Parsegian,Russel}. The resulting potentials for
two interplate separations, $h=1$ and $0.1$ $\mu$m, are shown in Fig.\
\ref{fig_numer}(a) along with the unconfined potential. The
interaction per volume squared has been scaled by $-T/R^6$. The
curves, therefore, represent the effective Hamaker coefficient
(divided by $\pi^2$, in units of $T$) as a function of interparticle
distance. The unconfined potential clearly exhibits the crossover from
the nonretarded $R^{-6}$ regime at small $R$ to the retarded $R^{-7}$
dependence at intermediate distances, and then back to the $R^{-6}$
decay due to temperature. The confined interaction deviates from the
unconfined one at distances $R\gtrsim h$. The amplification factor
increases moderately with distance until saturating to the asymptotic
value given in Eq.\ (\ref{Ninham99_conf}).
In the second example two polystyrene particles are confined in a
hydrocarbon slab of thickness $h=3$ nm, which is embedded in
water. This may mimic small hydrophobic inclusions in a biological
membrane. For the oily environment we used the permittivity
$\ein(i\xi)$ of pentane \cite{MN}. As is seen in Fig.\
\ref{fig_numer}(b), the amplification becomes significant only at
$R\gtrsim 10h$. This is because for smaller distances retardation has
not yet set in. (Note where the unconfined potential departs from its
nonretarded $R^{-6}$ behavior.)

The confinement-induced enhancement of the retarded dispersion
interaction can be viewed as a consequence of multiple-reflection
waveguiding of the electromagnetic radiation between the boundaries.
Despite the demonstrated strong effect the interaction remains weak.
In the example presented in Fig.\ \ref{fig_numer}(a), for instance,
the interaction free energy for $h=1$ $\mu$m and $R=10$ $\mu$m is
about $3\times 10^{-6}$ $T/\mu$m$^6$. Nevertheless, to achieve the
same energy without confinement one would have to set the
interparticle distance at about $2$ $\mu$m.  Thus, besides the
fundamental significance of the strong confinement effect reported
here for particle interactions in confined systems, it may become
useful also in extending the range of observation of the
Casimir-Polder interaction.

\begin{acknowledgments}
  We benefited from discussions with D.\ Andelman, H.\ Bary-Soroker,
  D.\ Bergman, S.\ Marcovitch, B.\ Ninham, A.\ Nitzan, S.\ Nussinov,
  A.\ Parsegian, R.\ Podgornik, and T.\ Witten. The work was supported
  by the Israel Science Foundation (77/03).  H.D.\ acknowledges the
  Israeli Council of Higher Education (Alon Fellowship).
\end{acknowledgments}




\begin{thebibliography}{99}
\bibitem{ft_static} Note that in the retarded limit the interaction is
  determined by the static ($\omega\rightarrow 0$) properties of the
  system.
\bibitem{ft_frequency} In fact, all the results depend only on the
  values of the dielectric functions along the imaginary axis,
  $\epsilon(i\xi)$ with real $\xi$ \cite{MN}.  These are real,
  positive, monotonously decreasing functions of $\xi$ involving the
  imaginary part of $\epsilon(\omega)$ alone.
\bibitem{ft_Tto0} The limit is taken such that
  $R/\lambda_T\rightarrow 0$ but $\lambda_0/\lambda_T$ is finite.  If
  $\lambda_0/\lambda_T\rightarrow 0$, the sum can be turned into an
  integral, which leads back to the zero-temperature results.
\bibitem{ft_salt} If the liquid contains mobile ions, however, the
  interaction will be screened beyond the wavelength corresponding to
  the ionic plasma frequency \cite{Russel}.
\end{thebibliography}
\end{document}